\documentstyle[epsf,pra,aps,preprint]{revtex}

\begin{document}
 
\title{Color-$SU(3)$-Ginzburg-Landau Effective Potential 
       for Order Parameter with ${\bf 3} \times {\bf 3}$ Symmetry}
\author{E. Nakano, T. Suzuki and H. Yabu }
\address{Department of Physics, Tokyo Metropolitan University, 
         1-1 Minami-Ohsawa, Hachioji, Tokyo 192-0397, Japan}
\date{\today}
\maketitle
\begin{abstract}
Ginzburg-Landau effective potential is studied 
for the order parameter that transforms 
in the $(3,3)$ representation 
under the color $SU(3)$ group.  
All the $SU(3)$ invariant terms 
to the fourth-order of the order parameter 
are classified and
the effective potential is constructed 
in its most general form. 
The conditions which stabilize the condensed phases are also obtained. 
As applications,  
the classification of the condensed phase is discussed in some special cases and 
the Higgs phenomenon associated with symmetry breaking 
is studied by introducing a coupling to the gauge boson.
\end{abstract}
%
%
\section{Introduction}
%
Recently, quark-pair condensation phenomena 
(color-superconductor)  
have received much interest 
in the high-density nuclear physics 
\makeatletter\cite{Barr,BaLo}\makeatother . 

In a high-density quark matter,  
quarks make a correlated Cooper pair 
due to the attractive one-gluon-exchange\cite{Iwa2,Wil1} or 
instanton-induced interactions\cite{Shur}. 
A lot of studies have been done for physical properties 
of color superconductors on the basis of the QCD-inspired 
models\cite{Son1,Wil1,Alf1,Gatto1,Berg}. 

From the symmetry point of view, 
the color-superconductor is characterized 
by the spontaneous symmetry breaking 
of the color  $SU_C(3)$ symmetry. 
Let's consider the quark field operator $q_\alpha(r)$  
with the color index $\alpha=1,2,3$ 
(other degrees of freedom, flavors and spins, 
are represented, e.g., as $q=u_{\uparrow}$). 
They transform as a vector (the ${\bf 3}$-representation)
under the $SU_C(3)$ transformation. 
(Irreducible representations of $SU(3)$ are denoted by their 
dimensions, e.g. {\bf 1}, {\bf 3}, {\bf 3}${}^*$, etc.)
The color-superconductor is a long-range ordered state, 
the order parameter of which is a two-body one:
$\Psi_{\alpha,\beta}(r) ={\langle q_\alpha(r) q'_\beta(r) \rangle}$. 
It is a second-rank tensor under $SU_C(3)$ with nine components, 
which makes a reducible ${\bf 3} \times {\bf 3}$-representation.   

Different from the systems with a scalar order parameter 
(metal superconductor, superfluid ${}^4$He, etc.), 
a variety of symmetry-breaking patterns shows up 
in the condensed states with the tensor order parameter. 
A typical example is the superfluid ${}^3$He, 
where, due to the p-wave nature of the Cooper pair, 
the order parameter has a tensor-type spin structure, 
$\Phi_{i,j}$ ($i,j =1,2,3$); 
it transforms as the ${\bf 3} \times {\bf 3}$-representation 
in the orbital and spin rotations 
$G =SO_L(3) \otimes SO_S(3)$ 
(also with the gauge $U(1)$ symmetry). 
Corresponding to several symmetry breaking patterns, 
many different condensed phases show up in the superfluid ${}^3$He: 
A-phase ($H =(S^2 \times SO(3))/Z_2$), 
B-phase ($R_B =S_1 \times SO_a(3)$), and so on. 
As discussed later, 
such a phase structure should exist also 
for the $SU(3)$-${\bf 3} \times {\bf 3}$-type order parameter, 
and it is very interesting to study the structures 
for such a novel symmetry breaking case\cite{Alf1}. 
(It should be noted that the ${\bf 3} \times {\bf 3}$ representation 
of $SU(3)$ is a complexification of that of $SO(3)$, 
so that the color superconducting state can be considered 
as a kind of the group extension for the superfluid ${}^3$He.)

The systems mentioned above are well described by
the thermodynamic effective free energy 
that is constructed as a function of the order parameter 
up to the fourth-order, i.e., 
Ginzburg-Landau method \cite{TT}.
This method incorporates the coherence nature of 
the pair condensation and
describes the second order phase transition.
It was also applied to the color-superconductor 
in some special cases \cite{Iida}.

In this paper, we discuss phase structures of 
the system with $SU(3)$-${\bf 3} \times {\bf 3}$ order parameter 
based on the Ginzburg-Landau method: 
the general form of the Ginzburg-Landau effective potential 
is presented to the fourth-order 
of the order parameter $\Phi_{\alpha,\beta}$, 
and analyzed in detail. 
As applications,  
the classification of the condensed phase and 
Higgs phenomenon are discussed in some special cases. 
%
\section{Structure of $SU(3)$-${\bf 3} \times {\bf 3}$ Order Parameter}
%
Let's consider the system with the symmetry group $SU(3)$, 
where the order parameter is given by the 3-by-3 matrix : 
$\Psi=(\Psi_{\alpha,\beta})$. 
For $g \in SU(3)$, it transforms as
\begin{equation}
     \Psi \longrightarrow g \Psi {}^tg, 
\label{eQaA}
\end{equation}
where ${}^tg$ is a transposition of $g$. 
It makes a ${\bf 3} \times {\bf 3}$-representation of $SU(3)$. 

The system should also be invariant 
under the $U(1)$ gauge transformation: 
\begin{equation}
     \Psi \longrightarrow e^{i\phi} \Psi. 
\label{eQaB}
\end{equation}

Because of the irreducible decomposition 
${\bf 3} \times {\bf 3} ={\bf 3}^* +{\bf 6}$, 
the order parameter $\Psi=(\Psi_{\alpha,\beta})$ 
can be decomposed into the irreducible components: 
$\Psi=S +A$, 
where $S$ and $A$ are the symmetric and antisymmetric 
components: 
\begin{equation}
     S ={1 \over 2} (\Psi +{}^t\Psi),  \quad
     A ={1 \over 2} (\Psi -{}^t\Psi).   
\label{eQa}
\end{equation}
Since the permutation symmetry for indices $(\alpha, \beta)$ are conserved under the 
$SU(3)$-transformation, 
the $S$ and $A$ become basis 
for the subrepresentations of $SU(3)$. 
By dimension counting, we find that 
the $S$ and $A$ correspond to the {\bf 6} and ${\bf 3}^*$ representations. 

From this consideration, 
symmetry breaking patterns for the condensed phases 
are classified into four types:
\begin{eqnarray}
\label{eQb}
     &\hbox{1) No-condensate}           &\quad A=S=0,     \label{con0}                 \\
     &\hbox{2) {\bf 3}${}^*$-condensate}&\quad A \neq 0, \quad S=0,  \label{con2}      \\
     &\hbox{3) {\bf 6}-condensate}      &\quad A=0,      \quad S \neq 0,  \label{con3} \\
     &\hbox{4) Mixed-condensate}        &\quad A \neq 0, \quad S \neq 0.  \label{con4}
\end{eqnarray}

Because of the $SU(3)$ invariance of the system, 
for $g \in SU(3)$, 
order parameters $(A,S)$ and $(g A {}^tg, g S {}^tg)$ 
represent the same condensed states, 
so that we can decompose the order parameters $A$ and $S$ ;
\begin{equation}
     A =e^{i\theta} G A_0 {}^tG,  \quad
     S =e^{i\theta} G S_0 {}^tG,
\label{eQc}
\end{equation}
where $G \in SU(3)$ and $e^{i\theta} \in U(1)$ 
are the Goldstone degrees of freedom which can be eliminated 
by the $SU(3)$ and $U(1)$ gauge transformations.  
Remaining parts $A_0$ and $S_0$ in eq.~(\ref{eQc}) 
determine essentially scales of the symmetry breaking. 
As proved in appendix~A, 
they can be parametrized as
\begin{equation}
     A_0 = \pmatrix{    0   &  \phi_3 & -\phi_2 \cr
                    -\phi_3 &     0   &  \phi_1 \cr
                     \phi_2 & -\phi_1 &     0   \cr},  \quad
     S_0 = \pmatrix{ d_1  & 0   &  0  \cr
                     0    & d_2 &  0  \cr
                     0    & 0   & d_3 \cr}, 
\label{eQd}
\end{equation}
where $d_{1,2,3}$ are real and $\phi_{1,2,3}$ are complex parameters: 
$\phi_i =|\phi_i| e^{i\theta_i}$. 
We take the parametrization in  eqs.~(\ref{eQc}) and (\ref{eQd}) 
as a standard form and call $d_i$ and $\phi_i$ 
normal coordinates of the ${\bf 3} \times {\bf 3}$ order parameter.
%
\section{Effective Potential for $SU(3)$-${\bf 3} \times {\bf 3}$ Order Parameter}
%
In the Ginzburg-Landau theory, 
the effective potential $V_{{\rm eff}}$ 
for the order parameter $\Psi$ plays a central role. 
It should be a invariant function of $\Psi$  
for the $SU_C(3)$ and $U(1)$ gauge transformations, 
(\ref{eQaA}) and (\ref{eQaB}). 
For a minimal Ginzburg-Landau model, 
it is necessary to construct $V_{{\rm eff}}$ 
to the fourth-order of $\Psi$.

The lowest order contributions to $V_{{\rm eff}}$ are the second-order 
terms in $\Psi$ and 
there exist such two terms constructed from $A$ and $S$: 
\begin{equation}
     V^{(2)}_{{\rm eff}} =a_1 {\rm Tr} A^* A +a_2 {\rm Tr} S^* S. 
\label{eQe}
\end{equation}
All other second-order terms can be shown to vanish 
or to be transformed into the above form 
(e.g., ${\rm Tr} A^* S =0$ due to the antisymmetry of $A^*S$.)

In a similar way, we can construct 
a general fourth-order potential $V^{(4)}_{{\rm eff}}$: 
\begin{eqnarray}
\label{eQf}
     V_{{\rm eff}}^{(4)} 
          &=&b_1 ({\rm Tr} A^*A)^2 
          +b_2 ({\rm Tr} S^*S)^2 +b_3 {\rm Tr} S^*SS^*S \nonumber\\
          &+&b_4 {\rm Tr} A^*A {\rm Tr} S^*S 
          +b_5 {\rm Tr} A^*A S^*S 
          +b_+  ({\rm Tr} A^*SA^*S +\hbox{c.c}) 
          +ib_- ({\rm Tr} A^*SA^*S -\hbox{c.c}). 
\label{eQea}
\end{eqnarray}
In the same way as for $V^{(2)}$, 
any other fourth-order terms can be transformed 
into the above form: e.g., 
$({\rm Tr} A^*A)^2 =2 {\rm Tr} A^*AA^*A$.

Using eq.~(\ref{eQc}), 
the potentials $V^{(2,4)}_{{\rm eff}}$, 
eqs.~(\ref{eQe}) and (\ref{eQf}),  
can be rewritten with the normal coordinates 
$(d_i,|\phi_i|,\theta_i)$ defined in eq.~(\ref{eQd}).  
For simplification, 
we introduce the variables 
$R \equiv \sqrt{|\phi_1|^2 +|\phi_2|^2 +|\phi_3|^2}$, 
$D \equiv \sqrt{d_1^2 +d_2^2 +d_3^2}$ 
and the normalized coordinates: 
$r_i \equiv |\phi_i|/R$ and $e_i \equiv d_i/D$, 
then 
eqs. (\ref{eQe}) and (\ref{eQf}) become
\begin{equation}
     V_{{\rm eff}}^{(2+4)} =g_R R^2 +g'_R R^4 
                           +g_D D^2 +(g'_D -h_D x) D^4
                           +(g''_D +h''_D u) R^2 D^2, 
\label{eQg}
\end{equation}
where the coefficients $g_i$ are defined by
linear combinations of $a_i$ and $b_i$ in eqs.~(\ref{eQe}) and (\ref{eQf}):
\begin{eqnarray}
     &&g_R =-2 a_1,  \quad
       g_R' =4 b_1,  \quad
       g_D =a_2,     \quad
       g_D'=b_2+b_3, \nonumber\\
     &&h_D =2 b_3,  \quad
     g''_D=-2 b_4 -b_5,  \quad
     h''_D=b_5. 
\label{eQi}
\end{eqnarray}
The variables $x$ and $u$ in eq.~(\ref{eQg}) are defined by 
\begin{equation}
     x \equiv e_1^2 e_2^2 +e_2^2 e_3^2 +e_3^2 e_1^2, 
\quad
     u \equiv f_1 r_1^2 +f_2 r_2^2 +f_3 r_3^2,     
\label{eQgA}
\end{equation}
where 
\begin{equation}
     f_1 =e_1^2 +c \sin(2\theta_1 +\delta) e_2 e_3,  \quad
     f_2 =e_2^2 +c \sin(2\theta_2 +\delta) e_3 e_1,  \quad
     f_3 =e_3^2 +c \sin(2\theta_3 +\delta) e_1 e_2. 
\label{eQgB}
\end{equation}
The constants $c$ and $\delta$ in eq.~(\ref{eQgB}) are obtained from 
the coefficients $b_\pm$ in eq.~(\ref{eQf}) as 
$c =-4\sqrt{b_+^2+b_-^2}/b_5$ and $\tan{\delta} =b_+/b_-$. 
%
\section{Domain of Variables}
%
Now 
we obtain the general form of the effective potential 
to the fourth-order, 
which is a function of the variables $(R,D,x,u)$. 
Let's consider the domain of these parameters. 

First, the variables $R$ and $D$ are defined as the lengths 
of $|\phi_i|$ and $d_i$, so that $R>0$ and $D>0$. 

We introduce new variables: 
\begin{equation}
     0 \leq p_i \equiv e_i^2 \leq 1,  \quad
     0 \leq q_i \equiv r_i^2 \leq 1,  \quad
     -1 \leq \lambda_i \equiv \sin(\theta_i +\delta) \leq 1,  \quad
     (i=1,2,3)
\label{eQj}
\end{equation}
which satisfies the constraints: $p_1+p_2+p_3=1$ and $q_1+q_2+q_3=1$. 
Geometrically, possible ranges 
in the $(p_1,p_2,p_3)$ or $(q_1,q_2,q_3)$ space 
become the two-dimensional region inside the equilateral triangle with 
the vertices $(1,0,0)$, $(0,1,0)$ and $(0,0,1)$.
Using these variables, 
the variables $x$ and $u$ are represented by
\begin{equation}
     x=p_2 p_3 +p_3 p_1 +p_2 p_3
      =\frac{1}{2} \left\{ 1-(p_1^2 +p_2^2 +p_3^2) \right\},  \quad
     u =f_1 q_1 +f_2 q_2 +f_3 q_3, 
\label{eQk}
\end{equation}
where $f_1 =p_1 +c \lambda_1 \sqrt{p_2 p_3}$. 

Let's consider the domain of $x =p_1 p_2 +p_2 p_3 +p_3 p_1$. 
It is clear that $x \geq 0$ because $p_{1,2,3} \geq 0$.  
Using the inequalities $(p_i+p_j)^2 \geq 4 p_i p_j$, 
we obtain 
\begin{equation}
     1 =(p_1+p_2+p_3)^2 
       =\frac{(p_1+p_2)^2 +(p_2+p_3)^2 +(p_3+p_1)^2}{2}
       +x \geq 
       \frac{4 x}{2} +x=3x. 
\label{eQkA}
\end{equation}
Thus, the variable $x$ takes the values $0 \leq x \leq 1/3$. 

The domain of $u =f_1 q_1 +f_2 q_2 +f_3 q_3$ 
depends on the value of $x$.  
Here, we summarize the derivation:
\begin{enumerate}
\item {\it Minimum/maximum of $u$ for fixed $f_i$.} 
Because $0 \leq q_i \leq 1$ with $q_1+q_2+q_3=1$, 
the domain of $u$ becomes \, 
$\min\{ f_1, f_2, f_3 \} \leq u \leq  \max\{ f_1, f_2, f_3 \}$. 
\item {\it Minimum/maximum of 
$f_1=p_1 +c \lambda_1 \sqrt{p_2 p_3}$ for fixed $p_i$.} 
Without loss of generality, we can assume $c \geq 0$ 
because $-1 \leq \lambda_1 \leq 1$.  
For fixed $p_i$, 
the variable $f_1$ takes a minimum value 
$p_1 -c \sqrt{p_2 p_3}$ at $\lambda_1=-1$ 
and a maximum value $p_1 +c \sqrt{p_2 p_3}$ at $\lambda_1=1$. 
Because $f_i$ are defined symmetrically for the index ($i=1,2,3$), 
similar results are obtained for $f_2$ and $f_3$. 
\item {\it Range of $u$ for fixed $p_i$.} 
Combining the above results 1. and 2. , 
we obtain the possible range of $u$ for fixed $p_i$: 
$f_1^- \leq u \leq f_1^+$, 
where
\begin{equation}
     f_1^- =p_1 -c \sqrt{p_2 p_3},  \quad
     f_1^+ =p_1 +c \sqrt{p_2 p_3}. 
\label{eQl}
\end{equation} 
\item Finally the domain of $u$ for the fixed value of $x$ 
is obtained by
$\min\left\{f_1^-,f_2^-,f_3^-\right\} \leq u \leq 
\max\left\{f_1^+,f_2^+,f_3^+\right\}$ 
where $\min$ and $\max$ are for $0 \leq p_i \leq 1$ 
with $p_1+p_2+p_3=1$. 
\end{enumerate}
Thus, the evaluation of the domain of $u$ 
has been reduced to those of $\min(f_-)$ and $\max(f_+)$,  
which are somewhat complicated because they depend not only 
on the coefficient $c$ but also on the variable $x$. 
The calculations for them are given in appendix~B, 
and we show only the final results for the domain of $(x,u)$ 
below.

The results are given for $(0 \le c \le 2)$ and $(2 \le c)$,
which are respectively shown in figure~1(a) and (b). 
These figures show upper and lower bounds of $u$ for fixed $x$.

\begin{eqnarray}
\hbox{a}: \ && 0 \le c \le 2 \nonumber \\ 
  &&  \hbox{Upper bound of}\ u : \frac{2-c}{3} \sqrt{1-3x} +\frac{1+c}{3} \label{ub1} \\ 
  &&  \hbox{Lower bound of}\ u : 
                                 \left\{ \begin{array}{ll}
                                       -c\sqrt{x} & (0 \le x \le \frac{1}{4}) \\
                                       -\frac{2+c}{3} \sqrt{1-3x} +\frac{1-c}{3} 
                                        & (\frac{1}{4} \le x \le \frac{1}{3})
                                       \end{array} \right. \label{lb1}
\end{eqnarray}
 The segment $\overline{AB}$ in figure~1(a) corresponds to eq.~(\ref{ub1}),
and $\overline{OC}$ and $\overline{CD}$ to two equations in (\ref{lb1}).
\begin{eqnarray}
\hbox{b}: \ && 2 \le c \nonumber \\
  &&  \hbox{Upper bound of}\ u : 
              \left\{ \begin{array}{ll}
                     -\frac{c-2}{3} \sqrt{1-3x} +\frac{1+c}{3} 
                      & \left( 0 \le x \le \frac{(2c -1)^2}{4(c^2 -c +1)^2} \right) \\
                     c\sqrt{x}
                      & \left( \frac{(2c -1)^2}{4(c^2 -c +1)^2} \le x \le \frac{1}{4} \right) \\
                     \frac{c-2}{3} \sqrt{1-3x} +\frac{1+c}{3}
                      & (\frac{1}{4} \le x \le \frac{1}{3}) 
                                    \end{array} \right. \label{ub2} \\
  &&  \hbox{Lower bound of}\ u : 
                               \left\{ \begin{array}{ll}
                                       -c\sqrt{x} & (0 \le x \le \frac{1}{4}) \\
                                       -\frac{2+c}{3} \sqrt{1-3x} +\frac{1-c}{3} 
                                        & (\frac{1}{4} \le x \le \frac{1}{3})
                                       \end{array} \right. \label{lb2}
\end{eqnarray}
 The segments $\overline{AE}$,$\overline{EF}$
and $\overline{FB}$ in figure~1(b) correspond to 
three divisions in eq.~(\ref{ub2}),
and $\overline{OC}$ and $\overline{CD}$ to two equations in (\ref{lb2}).
%
\section{Stability Condition}
%
For the system with the effective potential $V_{{\rm eff}}$ 
to be physically meaningful, 
it should be positive definite 
when the variables $R$ or $D$ take infinitely large value.  
(If not, any states with finite $R$ and $D$ 
become unstable.) 

This stability (positive definite) condition gives constraints 
on the parameters $g_i$ and $h_i$ in eq.~(\ref{eQg}) 
and we derive them in this section. 
First, we consider special cases:
\begin{enumerate}
\item When $D=0$ and $R \to \infty$, 
eq. (\ref{eQg}) become $V_{{\rm eff}} \sim g'_R R^4$
\item When $R=x=0$ and $D \to \infty$, 
we obtain  $V_{{\rm eff}} \sim g'_D D^4$. 
\end{enumerate}
From the positive definineness of $V_{{\rm eff}}$ 
in these limits, 
the first two constraints can be obtained:
\begin{equation}
     g'_D \geq 0,  \quad  g'_R \geq 0. 
\label{eQm}
\end{equation}
Hereafter, 
we consider the cases of $g'_D > 0$ and $g'_R > 0$ only. 
(We can treat the cases that $g'_D =0$ or $g'_R =0$  
with taking the limit of those in $g'_D > 0$ and $g'_R > 0$.) 

We replace the variables $R$ and $D$ by the scaled ones: 
$R \to R/(g'_R)^{\frac{1}{4}}$ and $D \to D/(g'_D)^{\frac{1}{4}}$, 
then eq.~(\ref{eQg}) becomes
\begin{equation}
     V_{{\rm eff}}=-G R^2 +R^4 -F D^2 +(1+F' x) D^4 
                   +2 (H +H' u) R^2 D^2, 
\label{eQn}
\end{equation}
where scaled parameters are defined by
\begin{equation}
     G =-\frac{g_R}{\sqrt{g'_R}},  \quad
     F =-\frac{g_D}{\sqrt{g'_D}},  \quad
     F'=-\frac{h_D}{g'_D},         \quad
     H =\frac{g''_D}{2\sqrt{g'_D g'_R}},  \quad
     H'=\frac{h''_D}{2\sqrt{g'_D g'_R}}. 
\label{eQo}
\end{equation}

For large values of $R^2$ and $D^2$, 
eq.~(\ref{eQn}) behaves as a linear quadratic form of them
for any fixed values of $(x,u)$: 
$V_{{\rm eff}} \sim R^4 +(1+F' x) D^4 +2(H+H' u) R^2 D^2$. 
The Sylvester's theorem should give the positive definite condition 
for it \cite{MIR}: 
\begin{eqnarray}
     &\hbox{a)}\ & 1+F' x -(H +H' u)^2 \ge 0 \nonumber \\
     && \quad \hbox{or}                    \nonumber \\
     &\hbox{b)}\ & 1 +F' x \ge 0 \quad\rm{and}\quad
                 H +H' u \ge 0,
\label{eQq}
\end{eqnarray} 
for any values of $(x,u)$, the domain of which have been shown
in figure~1(a) and (b).  
The second condition b) in eq.~(\ref{eQq}) should be required 
because $R^2$, $D^2$ $\geq 0$.

The condition (\ref{eQq}) gives constraints for the possible values 
of parameters $F'$, $H$ and $H'$,  
which can be obtained analytically. 
We sketch the derivation of them in Appendix~C,  
and only show the results here.
  
Stability conditions in $(H, F')$ space for fixed $c$ and $H'$ are classified 
at first into two types for ($0 \le c \le 2$) and ($2 \le c $),  
and then into four types by the values of $H'$ in each of the first two
(which correspond to four subscripts I$\sim$IV in figure~2 and figure~3), 
which are shown in figure~2(a)$\sim$(d) for $0 \le c \le 2$, 
                in figure~3(a)$\sim$(d) for $2 \le c $.
In the following lists, 
we obtain complete stability conditions for the coefficients 
$F'$, $H$, $H'$ and $c$ in eq.~(\ref{eQn}). 
%
\subsection{$0 \le c \le 2$ case}
%
 \begin{description}
   \item{a : } $H' \le -\frac{3}{2 -c}$ \ \ (figure~2(a))
  \begin{equation}
              F' \ge -3, \quad \ \ (-H'-1 \le H) \label{F2Ib}
  \end{equation} 
     This critical point $(-H'-1,-3)$ corresponds to 
     the point $A$ in figure~2(a).
  \item{b : } $-\frac{3}{2 -c} \le H^\prime \le 0$ \ \ (figure~2(b))
     \begin{eqnarray}
         F' &\ge k(c), 
          &\quad \quad \left( -H'-1 \le H \le -\frac{1 +c}{3}H' \right) 
              \quad \quad \overline{AB} \label{F2Id} \\
         F' &\ge -3, 
          &\quad \quad \left( -\frac{1 +c}{3}H' \le H \right) 
              \quad \quad \quad \quad \quad \quad \quad \overline{B\infty}  \label{F2Ie}
     \end{eqnarray} 
  where \ 
  \footnotesize{$k(c)=\frac{3}{2} \left[ 
                 \left( H +H' \right) \left( H +\frac{2c -1}{3}H' \right) -1
                 - \sqrt{ \left\{ (H +H')^2 -1 \right\} 
                          \left\{ \left( H +\frac{2c -1}{3}H' \right)^2-1 \right\} }  
                                                                      \right] $ }. 
    \normalsize  
  \item{c}:\ $0 \le H^\prime \le \frac{3}{2 +c}$ \ \ (figure~2(c))
   \begin{eqnarray}
          F' \ge& \frac{(c H^{\prime})^2}{1 -H^2}, \quad \quad \quad \quad \quad \ \  
           &(-1 \le H \le H_1) \quad \quad \quad \quad \quad \overline{\infty A} \label{F2Ig} \\
          F' \ge& (2H -cH^{\prime})^2 -4,\quad \ \ 
           &( H_1 \le H \le H_2(c) ) \quad \quad \quad \quad \overline{AB} \label{F2Ih} \\
          F' \ge& k(-c), \quad \quad \quad \quad \quad \quad \ 
            &\left( H_2(c) \le H \le \frac{c -1}{3}H^{\prime} \right) 
             \quad \overline{BC} \label{F2Ii} \\
          F' \ge& -3, \quad \quad \quad \quad \quad  \quad \quad \ \ 
            &\left( \frac{c -1}{3}H^{\prime} \le H \right) 
              \quad \quad \quad \quad \quad \overline{C\infty} \label{F2Ij}
   \end{eqnarray}
       where $H_1 \equiv \frac{cH' -\sqrt{16 +c^2 H'^2}}{4},\  
              H_2(c)=\frac{(3c +2)H'-\sqrt{16 +(c +2)^2 H'^2}}{4}$.
  \item{d}:\ $\frac{3}{2 +c} \le H'$ \ \ (figure~2(d))
   \begin{eqnarray}
         && F' \ge \frac{(c H')^2}{1 -H^2},\quad \quad \quad \quad \quad \ \ 
            (-1 \le H \le H_1 ) \quad \quad \quad \quad  \overline{\infty A} \label{F2Il} \\
         && F' \ge (2H -cH')^2 -4,\quad \ \ 
            \left( H_1 \le H \le \frac{cH'-1}{2} \right) 
            \quad \overline{AB} \label{F2Im} \\
         && F' \ge -3,\quad \quad \quad \quad \quad \quad \quad \ \ 
            \left( \frac{cH'-1}{2} \le H \right) 
            \quad \quad \quad \ \overline{B \infty} \label{F2In}
   \end{eqnarray}
%
\subsection{$2 \le c$ case}
%
 \begin{description}
  \item{a}: $H' \le -\frac{3}{c -2}$ \ \ (figure~3(a))
   \begin{eqnarray} 
       && F' \ge (2H +cH')^2 -4,\quad \ \ 
          \left( -H'-1 \le H \le -\frac{1 +cH'}{2} \right) 
          \quad \overline{AB} \label{F2IIb}\\
       && F' \ge -3,\quad \quad \quad \quad \quad \quad \quad \ \ 
         \left( -\frac{1 +cH'}{2} \le H \right)  
         \quad \quad \quad \quad \quad \ \ \overline{B \infty} \label{F2IIc}
   \end{eqnarray}
  \item{b}: $-\frac{3}{c -2} \le H' \le 0$ \ \ (figure~3(b))
   \begin{eqnarray} 
       && F' \ge (2H +cH^{\prime})^2 -4,\quad \ \ 
         (-H'-1 \le H \le H_2(-c)) \quad \quad \overline{AB} \label{F2IIe} \\
       && F' \ge k(c),\quad \quad \quad \quad \quad \quad \quad 
          \left( H_2(-c) \le H \le -\frac{1 +c}{3}H' \right) 
          \quad \overline{BC} \label{F2IIf} \\
       && F' \ge -3,\quad \quad \quad \quad \quad \quad \quad \ \,
          \left( -\frac{1 +c}{3}H' \le H \right)
          \quad \quad \quad \quad \quad \ \ \overline{C \infty} \label{F2IIg}
   \end{eqnarray}
  \item{c}: $0 \le H' \le \frac{3}{2 +c}$ \ \ (figure~3(c))
   \begin{eqnarray} 
       && F' \ge \frac{(c H')^2}{1 -H^2},\quad \quad \quad \quad \quad \ \ 
         (-1 \le H \le H_1)  \quad \quad \quad \quad \quad \overline{\infty A} \label{F2IIi} \\
       && F' \ge (2H -cH')^2 -4,\quad \ \ 
         ( H_1 \le H \le H_2(c))  \quad \quad \quad \quad \overline{AB} \label{F2IIj} \\
       && F' \ge k(-c),\quad \quad \quad \quad \quad \quad \           
         \left( H_2(c) \le H \le \frac{c -1}{3}H' \right) 
         \quad \overline{BC} \label{F2IIk} \\
       && F' \ge -3,\quad \quad \quad \quad \quad \quad \quad \ \ 
         \left( \frac{c -1}{3}H' \le H \right) 
         \quad \quad \quad \quad \quad \overline{C \infty} \label{F2IIl}
   \end{eqnarray}
  \item{d}: $\frac{3}{2 +c} \le H'$ \ \ (figure~3(d))
   \begin{eqnarray}
       && F' \ge \frac{(c H')^2}{1 -H^2},\quad \quad \quad \quad \quad \ \ 
          (-1 \le H \le H_1 ) \quad \quad \quad \ \ \overline{\infty A} \label{F2IIn} \\
       && F' \ge (2H -cH')^2 -4,\quad \ \ 
          \left( H_1 \le H \le \frac{cH'-1}{2} \right) \quad \overline{AB} \label{F2IIo} \\
       && F' \ge -3,\quad \quad \quad \quad \quad \quad \quad \ \ 
          \left( \frac{cH'-1}{2} \le H \right) 
          \quad \quad \quad \ \ \overline{B \infty} \label{F2IIp}
   \end{eqnarray}
 \end{description}
\end{description}
%
\section{Applications}
%
In this section we show some applications to our model.
As a first application,  
we obtain the phase diagrams for the two special sets of the coefficients (\ref{eQo}). 
As a second application,  
we construct a kinetic term by introducing a coupling to the gauge boson and 
study the Higgs phenomenon in the three characteristic phases.
%
\subsection{Phase diagrams for some cases}
%
\subsubsection{The angular parameter $(x,u)$ independent potential}
%
We study the condesed phases in the case of $(x,u)$-independent potential,   
$F'=0$ and $H'=0$ in eq.~(\ref{eQn}).
The effective potential becomes a function of $R$ and $D$
with three parameters $(G,F,H)$: 
\begin{equation}
V(R,D) = -G R^2 +R^4 -F D^2 +D^4 +2 H R^2 D^2. \label{eqV}
\end{equation}

The stability conditions are given as in the previous section: 
\begin{equation}
     1-H^2 \ge 0 \quad \hbox{or} \quad H \ge 0.  \label{eQu} 
\end{equation}

Four kinds of condensesd phase 
(as extremum points of the effective potential) are obtained: 
1) $R_0=D_0=0$ , \, 
2) $R_0 \neq 0$, $D_0=0$ , \, 
3) $R_0=0$, $D_0 \neq 0$ , \, 
4) $R_0 \neq 0$, $D_0 \neq 0$ 
corresponding to eqs.~(\ref{con0})-(\ref{con4}), 
and the parameter regions for each phase are shown in Table~1
with the conditions (\ref{eQu}).
Referring to Table~1, 
the phase diagram on $(F,\, G)$ space for fixed $H$ 
is obtained in figure~4
by comparing $V(R_0,\, D_0)$
in the ranges where more than two kinds of extrema exist.
At $H$$=$$-1$, 
the mixed phase emerges 
above the line $\overline{AB}$ ($G$$=$$-F$) in figure~4(a), 
and as $H$ increases to $1$
the two branches of this line 
merges into the line $\overline{OA}$ ($G$ $=$ $F$) in figure~4(c).
When $H$ exceeds $1$ the mixed phase vanishes.
%
\subsubsection{The potential for the symmetric phase}
%
Next
we study the case in which only the symmetric ${\bf 6}$ phase emerges, 
i.e.,  
$g_R$$=$$g'_R$$=$$g''_D$$=$$h''_D$$=$$0$ in eq. $(\ref{eQg})$.
In this case the effective potentail is given by
\begin{equation}
   V(x,D) = g_D D^2 +(g'_D-h_D x)D^4. \label{pot6}
\end{equation}
 From $0 \le x \le \frac{1}{3}$, 
we obtained the stability condition by 
\begin{eqnarray} 
   h_D \le 3\, g'_D, \ \ g'_D \ge 0. \label{eQp1}
\end{eqnarray}
This condition coincides with the previous results 
in $F$$\ge$$-3$ and $H'$$=$$H$$=0$, e.g., 
in eqs.~(\ref{F2Id}) and (\ref{F2Ie}).
From eq.~(\ref{pot6}) 
the four kinds of condensed phase 
are obtained: 
1) $D_0$$=$$0$ ,   $x$$=$$0$,\, 
2) $D_0$$\neq$$0$, $x$$=$$0$, \, 
3) $D_0$$\neq$$0$, \, 
4) $D_0$$\neq$$0$, $x$$=$$\frac{1}{3}$.
The parameter regions and 
the form of the order parameter $S_0$ in eq.~(\ref{eQd}) 
for each phase are given in Table~2. 
The phase diagram in ($F$, $G$) space for $g_D$$\ge$$0$ is shown in figure~5.

It is noticed that
the linear dependence of the effective potential (\ref{pot6}) on $x$ is 
the reason why the phase characterized by, e.g., $S_0$$=$$(d_1, d_2, 0)$ does not emerge. 
%
\subsection{Higgs phenomenon}
%
Higgs phenomenon is the one where
gauge bosons acquire finite effective masses
in the broken-symmetry phase.
In the minimal coupling scheme, 
the gauge boson fields $A^a_\mu$ are introduced through 
the kinetic terms of the order parameter.
To the second-order, 
it is given by
\begin{equation}
         K_{kinetic}
       = -k_A {\rm Tr} (D_\mu A)^* D^\mu A
         +k_S {\rm Tr} (D_\mu S)^* D^\mu S,         \label{eQz}
\end{equation}
where the coefficients $k_A$ and $k_S$ are
positive constants.
The covariant derivative $D_\mu$ in eq.~(\ref{eQz}) is defined by
\begin{equation}
      D_\mu \Psi \equiv 
      \partial_\mu \Psi -ig\frac{\lambda_a}{2}
      A^a_\mu \Psi  -ig \Psi \frac{{}^t\lambda_a}{2} A^a_\mu,       \label{eQx}
\end{equation}
where $g$ and $\lambda_{a=1 \sim 8}$ 
are the gauge coupling constant and
the Gell-Mann matrices \cite{LEE}.

The effective masses of the gauge bosons are obtained 
from the second-order terms of the gauge boson fields 
in the kinetic terms (\ref{eQz}).
%
\subsubsection{The ${\bf 3^*}$ condensed phase}
%
In the case of the ${\bf 3^*}$ condensed phase 
we consider an order parameter
\begin{equation}
       \Psi_0 =  \pmatrix{ 0 & -a & 0 \cr 
                           a &  0 & 0 \cr 
                           0 &  0 & 0 \cr },                   \label{eQQa}
\end{equation}
where $a$ is a real constant.
Note that this form is general enough
since other forms of the ${\bf 3^*}$ phase is obtained by 
a simple transformation to $\Psi_0$ as in eq.~(\ref{eQc}).
In this phase $SU_c(3)$ symmetry breaks into $SU(2)$.

By substituting $\Psi_0$ into eq.~(\ref{eQz}) 
the mass terms are obtained: 
$m_v^2$${\rm Tr}$$v^\dagger$$v$ and $m_\phi^2$${\rm Tr}$$\phi^2$ 
where 
$m_v$$\equiv$$\sqrt{2k_A}$$|a|$$g$, 
$m_\phi$$\equiv$$\sqrt{\frac{4}{3}k_A}$$|a|$$g$, 
$v$$=$$\frac{1}{2}$$(\lambda_4+i\lambda_5)$$(A_\mu^4+iA_\mu^5)$
$+$$\frac{1}{2}$$(\lambda_6+i\lambda_7)$$(A_\mu^6+iA_\mu^7)$ and 
$\phi$$=$$\lambda_8$$A_\mu^8$.
The gauge fields $v$ and $\phi$ 
are different multiplets of the residual symmetry $SU(2)$.

This phase is the one which is featured in most microscopic models.
Using the values of the gauge coupling constant $g$ and 
the gap $|a|$ caluculated in models \cite{Shur},
the effective masses are estimated: 
$m_v$$=$$\sqrt{\frac{3}{2}}$$m_\phi$\, $\sim$\, $100$ {\rm MeV}.
%
\subsubsection{The ${\bf 6}$ condensed phase}
%
We select an order parameter proportional to the identity matrix 
which is discussed in the previous application, see Table~2 :
\begin{equation}
     \Psi_0 = \pmatrix{ a & 0 & 0 \cr 0 & a & 0 \cr 0 & 0 & a \cr }. \label{eQQi}
\end{equation}
In this phase $SU(3)$ symmetry breaks into $SO(3)$.
Since the gauge boson field $A_\mu$($\equiv$$\lambda_aA_\mu^a$) is
transformed as direct products of two triplets 
under the $SO(3)$ group, 
they are decomposed into the direct sum of triplet (anti-symmetric part)
and quintet (symmetric part): 
\begin{equation}
      A_\mu = V_{\bf 3}+T_{\bf 5}, \ \ \ 
          (V_{\bf 3} \equiv \sum_{a=2,5,7} \lambda_a A^a_\mu), \ 
          (T_{\bf 5} \equiv \sum_{a=1,3,4,6,8} \lambda_a A^a_\mu). \label{eQQk}
\end{equation}
 Since $T_{\bf 5}$ is assigned to the generators of the broken symmetry $SU(3)/SO(3)$, 
the mass term is accompanied with the quintet term: 
     $m_T^2$${\rm Tr}$$(T_{\bf 5})^2$,  
     where $m_T$$\equiv$$\sqrt{4k_S}$$|a|$$g$. 
%
\subsubsection{The Mixed condensed phase}
%
We study a special case for the mixed phase 
characterized by the order parameter: 
\begin{equation}
     \Psi_0 = \pmatrix{ 0 & 0 & 0 \cr 0 & 0 & 0 \cr a & 0 & 0 \cr }. \label{eQQQi}
\end{equation}
In this phase $SU(3)$ symmetry breaks into $U(1)$ 
which is generated by $H(t)$$=$$e^{i(\lambda_3 +\sqrt{3}\lambda_8)t}$.
Since in this case the multiplets of the gauge field are not trivial, 
we investigate a dependence of the gauge field $A_\mu$ 
on the infinitesimal transformation of $H(t)$. 
\begin{equation}
      (H A_\mu H^\dagger)_{kl} \, \sim
        (A_\mu)_{kl}+ (P)_{kl}(A_\mu)_{kl},             \quad 
       P =i\pmatrix{  0 & 1 & 2  \cr  
                     -1 & 0 & 1  \cr 
                     -2 &-1 & 0  \cr}t          \label{eQQQp}
\end{equation} 
Noticing that 
 $(\lambda_3 +\sqrt{3} \lambda_8)$ and
 $(\lambda_3 -\frac{1}{\sqrt{3}} \lambda_8)$
are orthogonal each other,
the gauge field are decomposed into the multiplets: 
\begin{eqnarray}
    && A_\mu =\pmatrix{  0    & \xi     & \eta  \cr  
                  \xi^* &  0      & \zeta \cr 
                 \eta^* & \zeta^* & 0     \cr}
       +(\lambda_3 -\frac{1}{\sqrt{3}} \lambda_8) \varphi 
       +(\lambda_3 +\sqrt{3} \lambda_8) \varphi',              \label{eQQQr} \\
    &&  \xi \equiv A^1_\mu -iA^2_\mu,                                \quad 
        \eta \equiv A^4_\mu -iA^5_\mu,                               \quad  
        \zeta \equiv A^6_\mu -iA^7_\mu                                \nonumber \\ 
    &&    \varphi \equiv (3A^3_\mu -\sqrt{3}A^8_\mu)/4,             \quad
          \varphi' \equiv (A^3_\mu +\sqrt{3}A^8_\mu)/4.      \nonumber
\end{eqnarray}   
The mass terms are summerized as follows: 
\begin{eqnarray}
     && m_\xi^2 \, \xi^* \xi 
        =\frac{1}{2}( k_A +k_S) |a|^2 g^2 \xi^* \xi,                \quad \quad
        m_\eta^2 \eta^* \eta 
        =2 k_S |a|^2 g^2 \eta^* \eta                                 \nonumber \\
     && m_\zeta^2 \zeta^* \zeta 
        =\frac{1}{2}( k_A +k_S) |a|^2 g^2 \zeta^* \zeta,             \quad \quad
        m_\varphi \varphi^2 
        =\frac{16}{18}( k_A +k_S) |a|^2 g^2 \varphi^2.              \label{eQQQs}
\end{eqnarray}
In the same way more complicated cases can also be studied.
%
\section{Summary}
%
We have derived the general form of 
the Color-$SU(3)$-Ginzburg-Landau Effective Potential 
for the order parameter with ${\bf 3}$$\times$${\bf 3}$ symmetry 
to the fourth-order 
and determined the range of coefficients 
which stabilizes the vacuum of the system. 
Corresponding to the irreducible components of the order parameter
${\bf 3}$$\times$${\bf 3}$$=$${\bf 3^*}$$+$${\bf 6}$, 
we found four kinds of phases:
non condensed phase, 
anti-symmetric (${\bf 3^*}$) phase, 
symmetric (${\bf 6}$) phase and 
mixed (${\bf 3^*}$$+$${\bf 6}$) phase.

As applications, 
we obtained the phase diagrams in the two special cases and
studied the Higgs phenomenon in some characteristic phases 
by gauging this model.

The full classification of the broken vacuum 
are analytically possible 
including the angular parameter $x$ and $u$, 
and will be discussed in different paper. 

%
%
\appendix
%
\section{Normal Coordinate Decomposition}
%
In this appendix, 
we show the decomposition of the 3-by-3 symmetric matrix $S ={}^tS$ 
in eq.~(\ref{eQd}). 
The proof is a variation of that 
used in the group representation theory \cite{MUR}, 
the $SU(3)$-chiral nonlinear sigma model \cite{LEE} 
and the Kobayashi-Maskawa theory of quark mass matrix \cite{KM}.

We take the subsidery hermite matrix $H =S^* S =S^\dag S$. 
Its eigenvalue $\lambda_i$ and eigenvectors ${\bf e}_i$  
are obtained from the eigenequation:
$H {\bf e}_i =\lambda_i {\bf e}_i$  ($i=1,2,3$). 

In this appendix, we consider only the case of no degeneracy:
$\lambda_i \neq \lambda_j$ ($i \neq j$). 
(The extension to degenerate cases should be made
by analytic continuation. )
 
Because of the hermiticity of $H$, 
the eigenvectors ${\bf e}_i$ can be taken 
to be orthogonal unit vectors
(${}^\dag{\bf e}_i{\bf e}_j =\delta_{ij}$), 
and the eigenvalues $\lambda_i$ take real positive values. 
(because of $H=S^\dag S$). 
In matrix form, the eigenequation becomes
\begin{equation}
     U^\dagger H U =L 
          \equiv \pmatrix{\lambda_1 & 0 & 0 \cr
                          0 & \lambda_2 & 0 \cr
                          0 & 0 &\lambda_3  \cr}, 
\label{eQaPaA}
\end{equation}
where $U =({\bf e}_1, {\bf e}_2, {\bf e}_3) \in SU(3)$.

The eigenvector ${\bf e}_i$ satisfies 
\begin{equation}
     H^* S {\bf e}_i =S S^* S {\bf e}_i 
                    =S H {\bf e}_i 
                    =\lambda_i S {\bf e}_i.  
\label{eQaPb}
\end{equation} 
so that we obtain $S {\bf e}_i =\alpha_i {\bf e}^*_i$ 
(because $H^* {\bf e}^*_i =\lambda_i {\bf e}^*_i$).  
$\alpha_i \equiv |\alpha_i| e^{i\theta_i}$ are complex numbers. 
In matrix form, it becomes
\begin{equation}
     {}^tU S U =Z 
          \equiv \pmatrix{\alpha_1 & 0 & 0 \cr
                          0 & \alpha_2 & 0 \cr
                          0 & 0 & \alpha_3 \cr}. 
\label{eQaPd}
\end{equation}
The polar decomposition of $Z$ becomes 
\begin{equation}
     A =R e^{iT} 
       =\pmatrix{|\alpha_1| & 0 & 0 \cr
                 0 & |\alpha_2| & 0 \cr
                 0 & 0 & |\alpha_3| \cr} 
     \pmatrix{ e^{i\theta_1} & 0 & 0 \cr
               0 & e^{i\theta_2} & 0 \cr
               0 & 0 & e^{i\theta_3} \cr},  \quad
     T =\pmatrix{ \theta_1 & 0 & 0 \cr
                  0 & \theta_2 & 0 \cr
                  0 & 0 & \theta_3 \cr}. 
\label{eQaPe}
\end{equation}

Because of
$\lambda_i {\bf e}_i =H {\bf e}_i =S^* S {\bf e}_i 
                     =\alpha_i S^* {\bf e}^*_i 
             =\alpha_i (S {\bf e}_i)
             =|\alpha_i|^2 {\bf e}_i$, 
we obtain $\lambda_i =|\alpha|_i$ and also $R =\sqrt{L}$. 
The polar decomposition of $A$ becomes $A=\sqrt{L} e^{iT}$. 

Because the angular part $T$ is diagonal, 
it can be expanded by the 3-by-3 unit matrix $I$ 
and the diagonal Gell-Mann matrices $\lambda_{3,8}$:  
$T =\theta I +\phi_3 \lambda_3 +\phi_8 \lambda_8$. 
If we define $K =e^{\frac{i}{2}(\phi_3 \lambda_3 +\phi_8 \lambda_8)}$, 
then we obtain $e^{iT} =K e^{i\theta} I K$. 
Using the first equation in (\ref{eQaPe}), 
the matrix $S$ becomes
\begin{equation}
     S =U^* Z U =U^* \sqrt{L} e^{iT} U^\dagger 
                =e^{i\theta} U^* K \sqrt{L} K U^\dagger 
                =e^{i\theta} G \sqrt{L}\, {}^tG, 
\label{eQaPf}
\end{equation}
where $G \equiv U^* K$ and ${}^tG ={}^tK U^\dagger =K U^\dagger$. 
This is just the decomposition of $S$ in eq.~(\ref{eQc}) 
with $S_0 =\sqrt{L}$. 

If we define the matrix $A_0$ by $A =G A_0 {}^tG$, 
then $A_0$ also becomes antisymmetric 
and is generally represented 
as the first equation in eq.~(\ref{eQd}). 
%
\section{Maximum and Minimum of \lowercase{$f$}}
%
Instead of $p_i$ with $p_1+p_2+p_3=1$, 
we take independent variables $(s,t)$: 
\begin{equation}
     s =\frac{1}{\sqrt{2}} (-p_1+p_2),  \quad
     t =-\sqrt{\frac{3}{2}} \left( p_1+p_2-\frac{2}{3} \right). 
\label{eQaQa}
\end{equation}
The possible range of $(s,t)$ is determined from the condition: 
$0 \leq p_i \leq 1$ and $p_1+p_2+p_3=1$.  
Using (\ref{eQaQa}), 
it is found to be the inside of the equilateral triangle  
with vertices 
$(0,\sqrt{2/3})$ and $(\pm 1/\sqrt{2},-1/\sqrt{6})$
in the $(s,t)$ plane. 
Since the function $f(s,t)$ is an even function of $s$, 
$f(s,t) =f(-s,t)$, 
we may consider a half ($s \geq 0$) of the triangular region (figure~6) 
to find its maximum/minimum. 

For a fixed value of $x$, 
a trajectory of $x =p_2 p_3 +p_3 p_1 +p_1 p_2 =-(s^2+t^2)/2 +1/3$ 
in $(s,t)$-plane become a circle 
with the center $(0,0)$ and the radius $\sqrt{(2/3)-2x}$ (figure~6). 
From the aspect of intersections between the circle and the triangular region, 
two cases should be discriminated: 
\begin{enumerate}
\item  
When $0 \leq x < 1/4$, 
the circle intersects with three sides of the triangle 
(figure~6(a)), 
and the two separate arcs ($B_0B_1$ and $B_2B_3$ in figure~6(a))  
are included in the half triangular region.  
\item
When $1/4 \leq x \leq 1/3$, 
the whole circle is included in the triangular region 
(figure~6(b)). 
The intersections with the $s$-axis ($t=0$) are denoted by $B_0$ and $B_4$. 
\end{enumerate}
The $(s,t)$-coordinates of the above four points $B_{0\sim 4}$ are given by 
\begin{eqnarray}
     B_0 &=&\left( 0, \sqrt{\frac{2}{3}-2x} 
            \right),                                \quad
     B_1  = \left( \frac{1-\sqrt{1-4x}}{2\sqrt{2}}, 
                   \frac{1+3\sqrt{1-4x}}{2\sqrt{6}} 
            \right),                                \\
     B_2 &=&\left( \frac{1+\sqrt{1-4x}}{2\sqrt{2}}, 
                   \frac{1-3\sqrt{1-4x}}{2\sqrt{6}} 
            \right),                                \quad
     B_3  = \left( \sqrt{\frac{1-4x}{2}}, 
                  -\frac{1}{\sqrt{6}} 
            \right),                               \\
     B_4 &=&\left( 0, -\sqrt{\frac{2}{3}-2x} 
            \right). 
\label{eQaQb}
\end{eqnarray}

Let's consider the maximum value of $f_3^+$ 
for fixed values of $c$ and $x$.  
Using $s =\sqrt{6 -9 t^2 -18 x}/3$ and (\ref{eQaQb}), 
$f^+_3$ in eq.~(\ref{eQl}) becomes 
\begin{equation}
     f_3^+(t) =\frac{1}{3} \left[ \sqrt{6} t +1 
                               +c \sqrt{6 t^2 -\sqrt{6} t +9x -2} 
                           \right], 
\label{eQaQc}
\end{equation}
where we take $t$ as an independent variable 
and $(x,c)$ as any fixed parameters. 
Solving the equation $[f_3^+(t)]' =0$, 
we obtain one local extremum at 
$B_M =(\sqrt{6 -9 t_M^2 -18 x}/3,t_M)$ where 
\begin{equation}
     t_M =\frac{1}{2\sqrt{6}} 
          \left[ 1 -3\frac{ \sqrt{(4x-1)(c^2-1)} }{1-c^2} \right]. 
\label{eQaDd}
\end{equation}
for some values of $(x,c)$. 
The parameter range 
where the extremum of $f^+_3$ exists 
(at $B_M$) can be read off 
from the condition that 
$t_M$ in (\ref{eQaDd}) should take a real value:
i.e., $(4x-1)(c^2-1) \geq 0$. 
Solving this inequality, 
we obtain two separated regions: 
a) $c \leq 1$ \& $x \leq 1/4$ and 
b) $c \geq 1$ \& $x \geq 1/4$ 
(Shaded regions in figure~7) 
where the function $f^+_3$ has an extremum. 
Judging from the signature of $[f^+_3]''(t_M)$, 
we can find that $f^+_3(t_M)$ is a local maximum 
in region a) and minimum in region b). 
For $t_M$ have to be in the permissible region 
shown in figures~6(a) and (b), 
we obtain further condition: 
\begin{enumerate} 
\item
When $x \leq 1/4$ and $c \leq 1$, 
$t_M$ have to be on the arc $B_2B_3$ (figure~6(a)).  
Thus, we obtain $x \geq c^2/4$ ($\overline{OD}$ in figure~7). 
\item
When $x >1/4$, 
$t_M$ have to be on the half circle $B_0B_4$ (figure~6(b)). 
Thus, we obtain $c \leq 2$ or $x \leq c(5c-4)/[4(2c-1)^2]$ 
($\overline{AB}$ in figure~7),
\end{enumerate}
Summarizing the above discussions 
about the aspects of extrema of $f^+_3$,  
we can classify the five regions in the $(c,x)$-plane 
(figure~7). 

Let's analyze maximum points in each regions.
Candidates of the absolute maximum point of $f_+^3$ 
are the end points $B_{0\sim 4}$ where 
\begin{eqnarray}
     & f_+^3(B_0) = \frac{1+c}{3} 
                  +\frac{2-c}{3} \sqrt{1-3x}, \quad
     & f_+^3(B_1) = \frac{1}{2} 
                  +\frac{1}{2} \sqrt{1-4x},   \nonumber \\
     & f_+^3(B_2) = \frac{1}{2} 
                  -\frac{1}{2} \sqrt{1-4x},   \quad
     & f_+^3(B_3) =c \sqrt{x},                \nonumber \\
     & f_+^3(B_4) = \frac{1+c}{3} 
                  -\frac{2-c}{3} \sqrt{1-3x},
\label{eQaDe}
\end{eqnarray}
or the local maximum at $M$ (if exist):
\begin{equation}
     f_+^3(B_M) =\frac{1}{2} 
                -\frac{1}{2} \sqrt{(c^2-1)(4x-1)}.
\label{eQaDf}
\end{equation}
Comparing $f_+^3(B_i)$ in eqs. (\ref{eQaDe}) and (\ref{eQaDf}), 
we can find maximum points in each region in figure~7: 
\begin{enumerate}
\item The maximum is at $B_0$ in regions 1, 2, a part of 3 
($c\leq 2$ or 
 $x \leq (2c-1)^2/[4(1-c+c^2)^2]$ $\overline{EF}$ in figure~8 ),  
and a part of 4 ($c \leq 2$). 
\item The maximum is at $B_3$ in a remaining part of region 3. 
\item The maximum is at $B_4$ in region 5 and a remaining part of 4. 
\end{enumerate}
The results are summarized in figure~8. 

In a similar way, the minimum value of $f_3^-(t)$ can be calculated: 
\begin{equation}
     f_3^-(t) =\left\{ \begin{array}{ll}
                      c \sqrt{x} & ( 0 \leq x < 1/4) \nonumber\\
                      \frac{1-c}{3} -\frac{2+c}{9} \sqrt{1-3x}
                                 & (1/4 \leq x \leq 1/3) \nonumber
                       \end{array}\right.
\label{eQaDk}
\end{equation}
%
\section{Derivation of stability conditions}
%
In this appendix, 
we sketche the derivation of the stability condition: 
a parameter range of $(F',H,H')$ that satisfies (\ref{eQq}). 

Instead of $(x,u)$, 
we take new variables $(X,U)$ defined by
\begin{equation}
     X \equiv 1+F' x,  \quad 
     U \equiv H +H' u.  
\label{eQaFa}
\end{equation}
Using them, 
the conditions in (\ref{eQq}) become simple: 
\begin{equation}
     \hbox{a) }\, X \geq U^2, 
          \quad {\rm or}\quad
     \hbox{b) }\, X \geq 0 \ {\rm and}\ U \geq 0.  
\label{eQaFb}
\end{equation}
In the $(X,U)$-plane, 
the range that satisfies these conditions is represented by 
the hatched area in figure~9. 
We denote it by the region $\Sigma$. 

The possible range of variables $(X,U)$, 
which is translated from the one of $(x,u)$ (given in figure~1), 
depends on parameters $(F',H',c)$ 
and we denote it by $\Lambda(F',H',c)$. 
In figure~9, as an example, 
we show the $\Lambda(F',H',c)$ 
as the cross-hatched region 
for some values of $(F',H',c)$ that satisfies 
$F' \geq 0$, $H' \geq 0$ and $c \geq 2$. 
In $(X,U)$-plane, the stability condition (\ref{eQq}) 
become equivalent with the geometrical condition:
\begin{equation}
     \Lambda(F',H',c) \subset \Sigma. 
\label{eQaFc}
\end{equation}
The boundary  of $\Lambda(F',H',c)$ and $\Sigma$ 
are patchworks of line or parabolic segments, 
so that the condition (\ref{eQaFc}) can be solved 
algebraically.  

As an example, let's consider the case that  
$c \leq 2$, $F' \geq 0$ and $H' \geq 0$ (figure~9). 
In this case, eq. (\ref{eQaFc}) can be attributed 
into whether the boundary segment $\overline{\infty GO}$ for $\Sigma$ 
intersects that of $\Lambda$, AB, or not. 
Both segments, $\overline{\infty GO}$ and $\overline{AB}$, 
have parabolic shapes represented by 
$X =U^2$ ($\overline{\infty GO}$) and 
$X =[F'/(cH')^2] (U-H)^2+1$ ($\overline{AB}$). 
It can be judged from 
the condition that the edge points, 
$A=(1,H)$ and $B=(\frac{F'}{4}+1,H-\frac{cH'}{2})$,  
are included in $\Sigma$;
it gives two necessary conditions for parameters:
\begin{equation}
     -1 \leq H,  \quad
     \left( H -\frac{cH'}{2} \right)^2 \leq \frac{F'}{4}+1. 
\label{eQaFd}
\end{equation}
In case that $F'/(cH')^2 \leq 1$, 
the slope of $\overline{\infty GO}$ is more gradual than that of $AB$, 
and eqs. (\ref{eQaFd}) are sufficient condition.  
However, in the case that $F'/(cH')^2 \geq 1$, 
an extra condition should be added 
in order that two parabolic segments have no intersections: 
\begin{equation}
     (cH')^2 +F'(H^2-1) < 0.
\label{eQaFe}
\end{equation}
Eqs. (\ref{eQaFd}) and (\ref{eQaFe}) 
give the necessary and sufficient stability condition in that case:
\begin{equation}
     F' \geq \frac{(c H')^2}{1-H^2},  \quad
     F' \geq (2H -c H')^2 -4,         \quad
     -1 \le H, 
\label{eQaFf}
\end{equation}
which is shown as the hatched region in figure~2. 

Other cases can be treated in a similar way. 
%
%

%
%
\newpage
\begin{description}
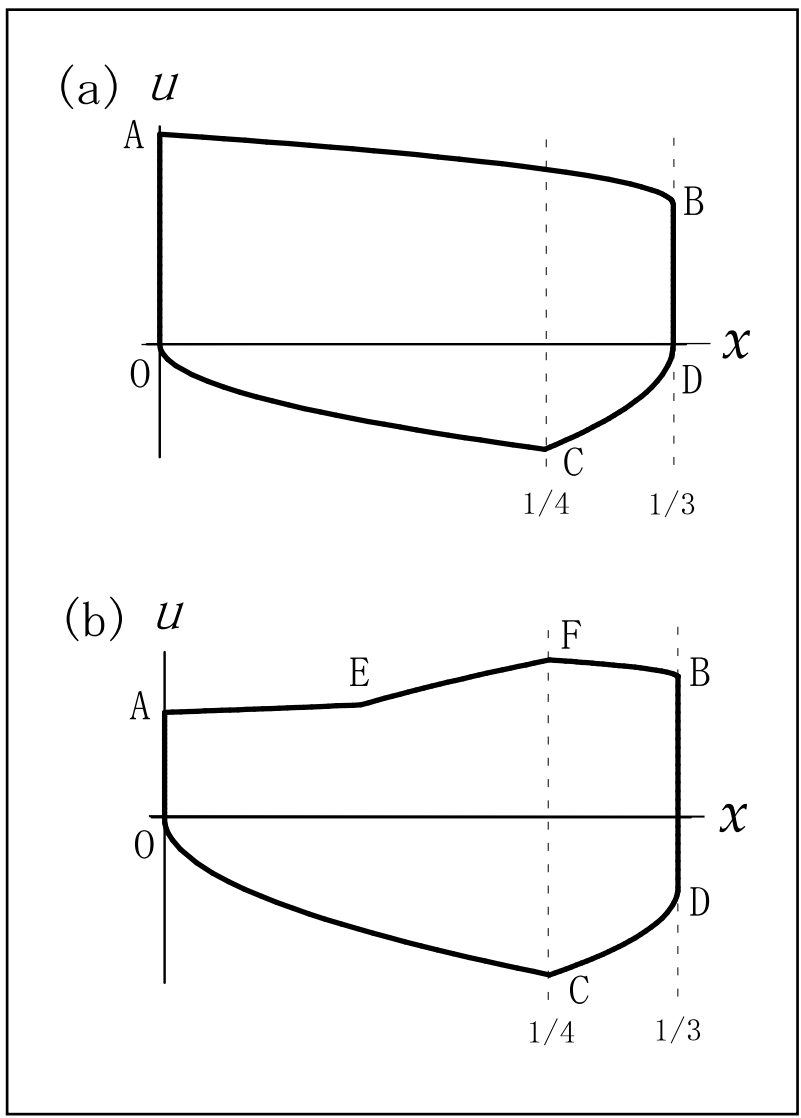
\item{figure~1.} 
Range of variables $x$ and $u$. 
a) for $0\le c \le 2$, \  
b) for $2\le c $. 
Details are given in the text.
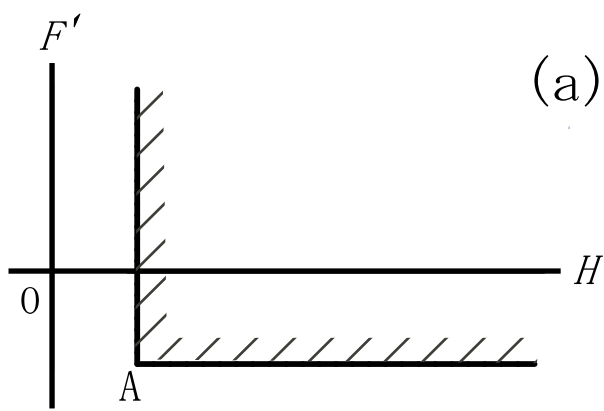
\item{figure~2.}
Stability region in $(H,F')$ space for $(0 \le c \le 2)$ 
(the shaded area).
     figure~2(a) is for $H'$ $\le$ $-\frac{3}{2-c}$, \,  
           (b)    for $-\frac{3}{2-c}$ $\le$ $H'$ $\le$ $0$, \, 
           (c)    for $0$ $\le$ $H'$ $\le$ $\frac{3}{2+c}$ \, 
      and  (d)    for $\frac{3}{2+c}$ $\le$ $H'$.
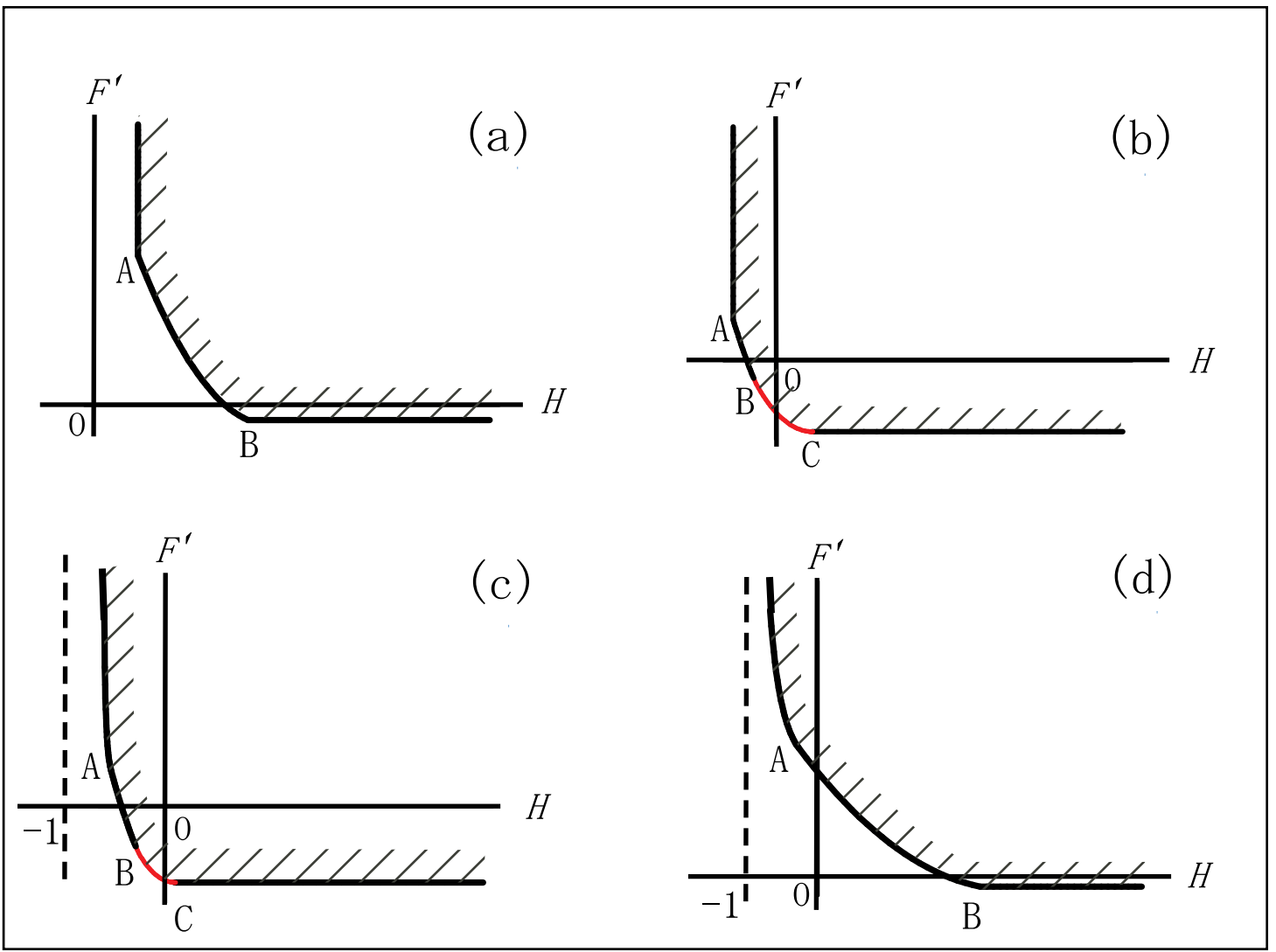
\item{figure~3.}
Stability region in $(H,F')$ space for $(2 \le c)$.
     figure~3(a) is for $H'$ $\le$ $-\frac{3}{c-2}$, \, 
           (b)    for $-\frac{3}{c-2}$ $\le$ $H'$ $\le$ $0$, \, 
           (c)    for $0$ $\le$ $H'$ $\le$ $\frac{3}{2+c}$ \,  
      and  (d)    for $\frac{3}{2+c}$ $\le$ $H'$.
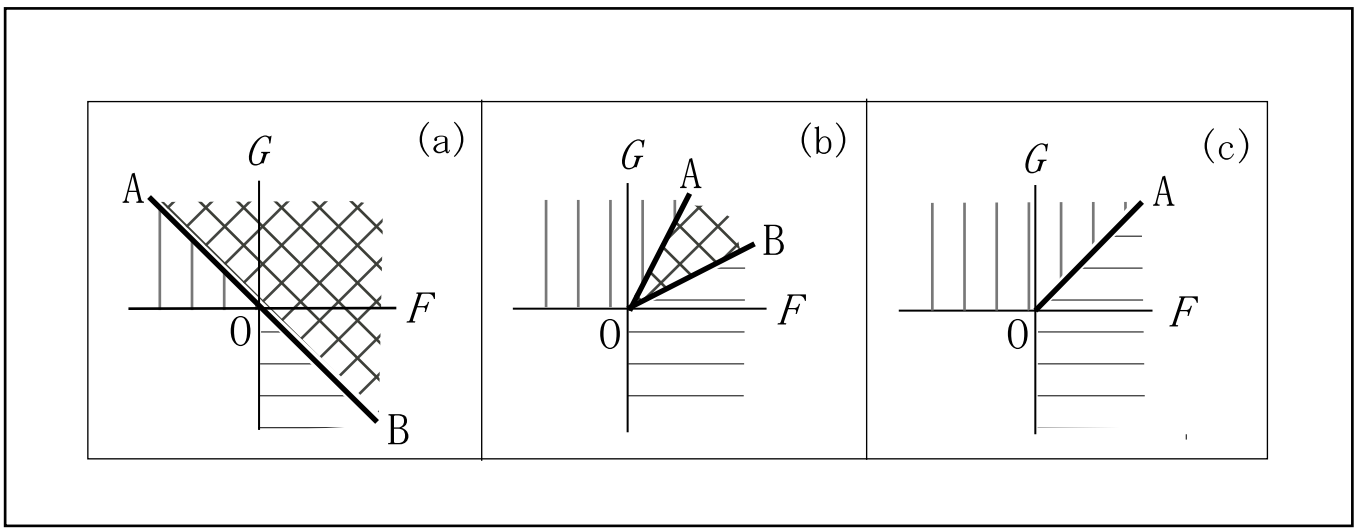
\item{figure~4.}
(a)  
The phase diagram for the effective potential (\ref{eqV}) 
in $(F, G)$ space when $H$$=$$-1$. $\overline{AB}$ : $G$$=$$-F$.
The regions with vertical stripes, 
with horizontal stripes and 
with cross hatch show ${\bf 3^*}$, ${\bf 6}$ and the Mixed condensed phases.
The remaining part ($G \le 0$ and $F \le 0$) shows the Non condensed phase.
(b) 
The same, but $H$$=$$1/2$.
$\overline{OA}$ : $G$$=$$2F$. 
$\overline{OB}$ : $G$$=$$F/2$.
(c) 
The same, but $H$$=$$1$.
$\overline{OA}$ : $G$$=$$F$.
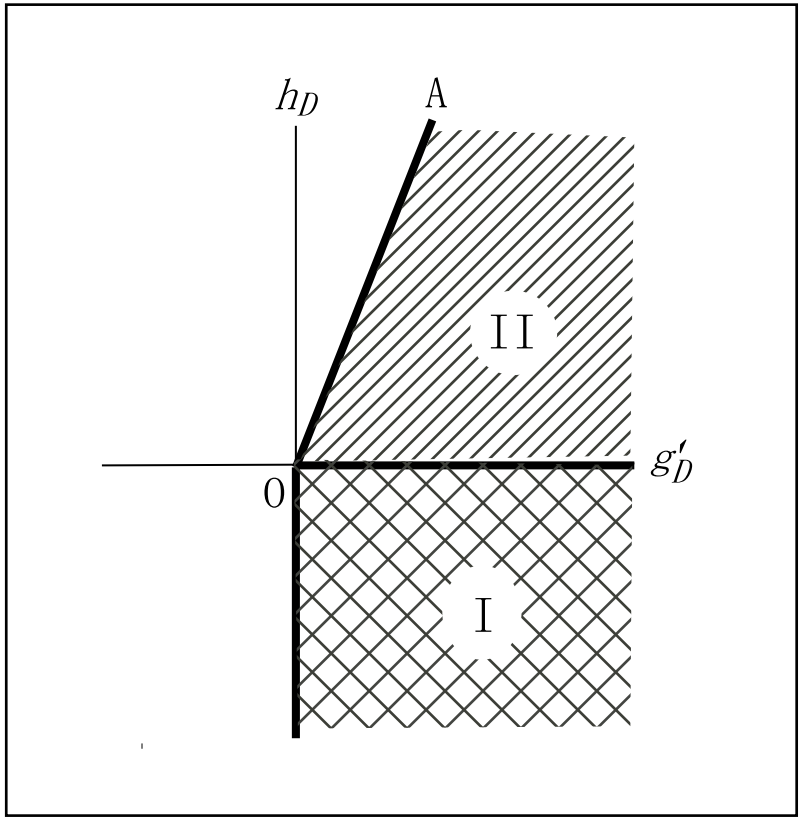
\item{figure~5.}
The phase diagram for the effective potential (\ref{pot6}) 
in $(g'_D, h_D)$ space when $g_D$$<$$0$. 
$\overline{AB}$ : $h_D$$=$$3g'_D$.
The regions I and II corresponds to the phases
2) and 4) in Table~2, 
but 
the region just on the line $h_D$$=$$0$ ($g'_D$$>$$0$) corresponds to the phase 3).
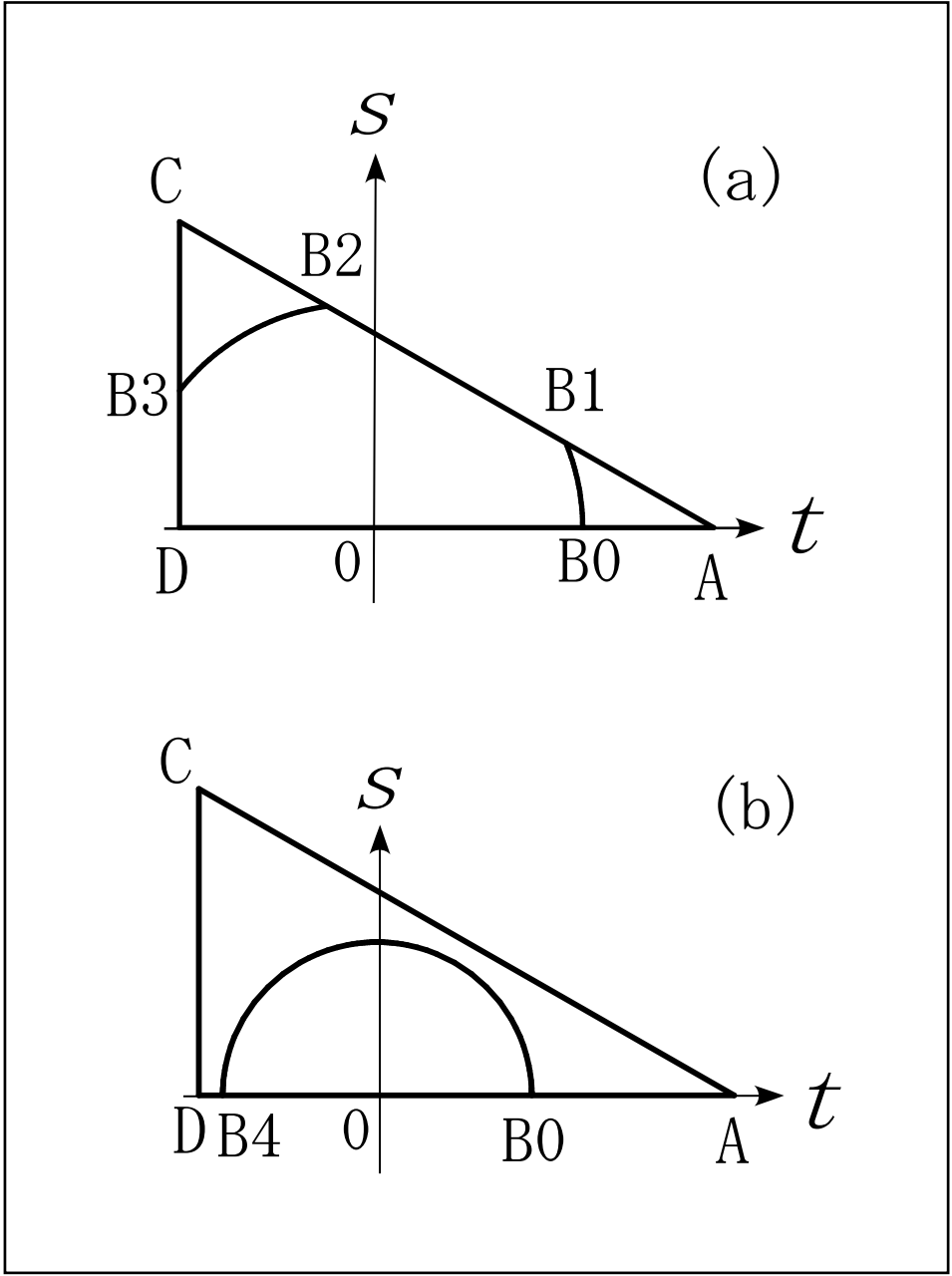
\item{figure~6.}
Range of variables $s$ and $t$ is shown as 
inside of the half triangle ACD.
Trajectories of ($s$, $t$) for fixed $x$ are given 
for $0 \le$ $x$ $\le \frac{1}{4}$
by two separate arcs $B_0 B_1$ and $B_2 B_3$
in (a)  
and 
for $\frac{1}{4} \le x \le \frac{1}{3}$
by a half circle $B_0 B_4$
in (b). 
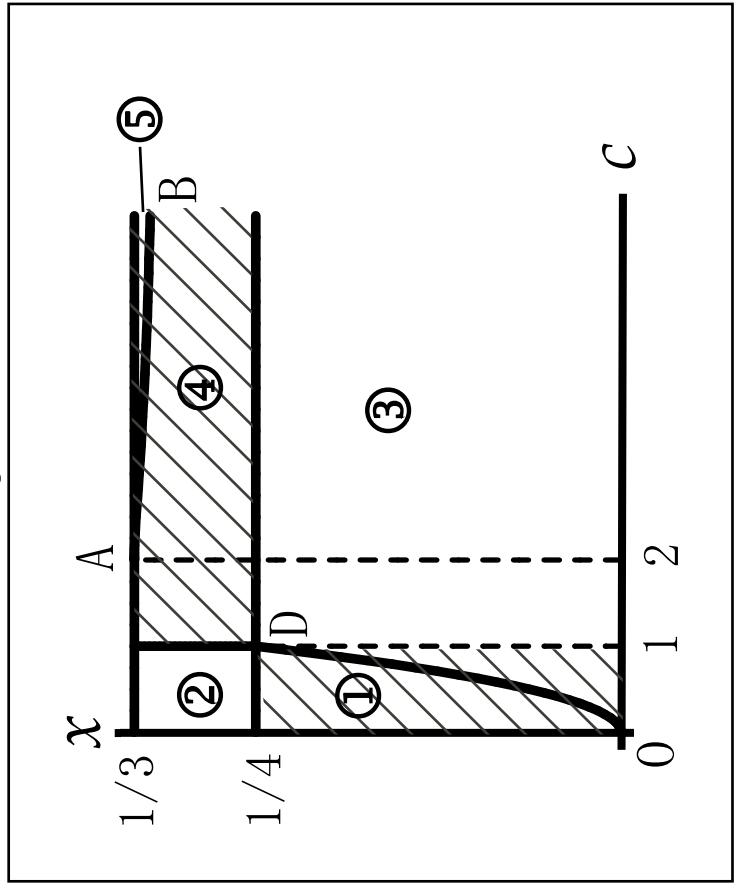
\item{figure~7.}
The five regions in the $(c,x)$-plane according to 
the aspects of $t_M$. 
$t_M$ is defined in the shaded regions.
In addition, 
in the region 1 and 4 
$t_M$ exists on the arc $B_2 B_3$ (figure~6(a)), 
and on the half circle $B_0 B_4$ (figure~6(b)).
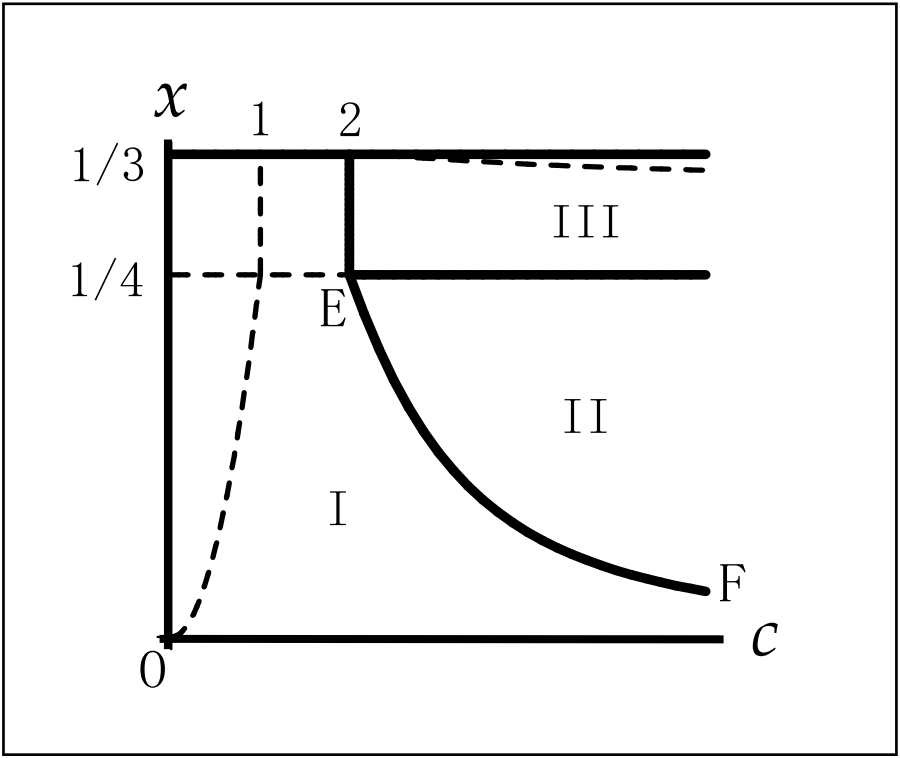
\item{figure~8.}
The three regions in the $(c,x)$-plane separated by
the maximum values of $f^3_+$,
which are 
$f^3_+(B_0)$ , $f^3_+(B_3)$ and $f^3_+(B_4)$
in the region I, II and III.
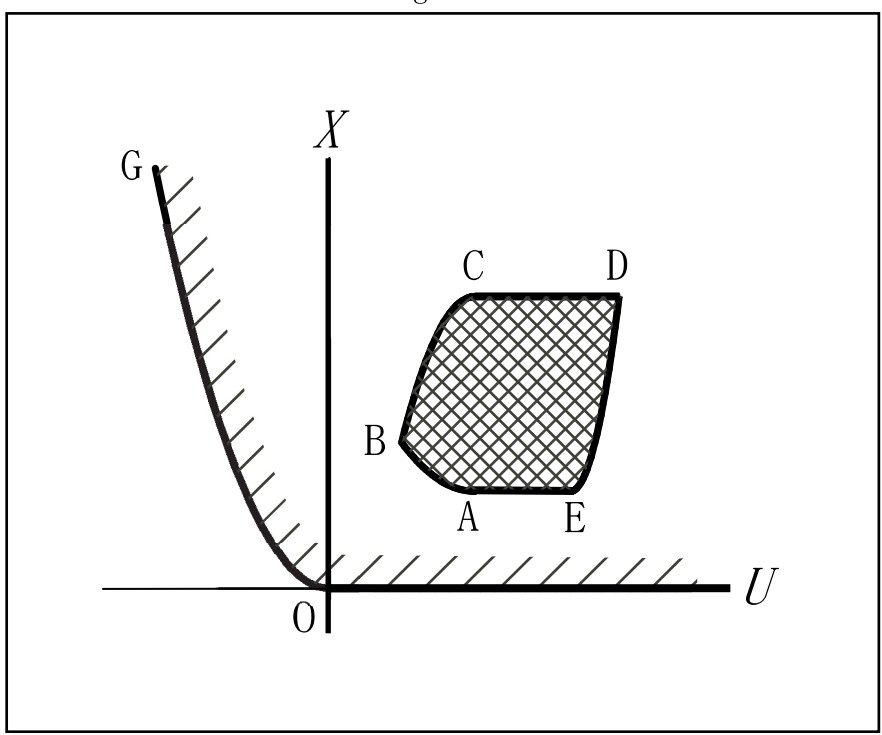
\item{figure~9.}
An illustration for the stability condition $\Lambda(F',H',c) \subset \Sigma$ 
in the case $c$ $\le$ $2$, $F'$ $\ge$ $0$ and $H'$ $\ge$ $0$.
\end{description}
\newpage
\begin{center}
Table~1 \ :\ 
Condensed phases and their parameter ranges. \\
\end{center}
\begin{center}
     \begin{tabular}{| c | c | c | c | c |} \hline
           & $R_0^2$ & $D_0^2$ & $V(R_0,\, D_0)$ 
           & the ranges where $R_0,\, D_0$ are defined \\ \hline 
     ${\rm Non\, cond.}$ 
           & $0$ & $0$ & $0$ 
           & $G \le 0,\quad F \le 0$                       \\
     ${\bf 3^*}\, \hbox{cond.}$ 
           & $G/2$ & $0$ & $ -G^2/4$  
           & $G \ge 0$                                     \\
     ${\bf 6}\, \hbox{cond.}$ 
           & $0$ & $ F/2$ & $-F^2/4$  
           & $F \ge 0$                                     \\
     ${\rm Mix\, cond.}$ 
           & $\frac{G -H F}{2 (1 -H^2)}$ & $\frac{F -H G}{2(1 -H^2)}$ 
           & $\frac{-G^2 -F^2 +2 F G H}{4(1 -H^2)}$ 
           & ${} \left\{  \begin{array}{ll}
                            G \ge H F \, \hbox{and} \, G \ge F/H, 
                         & {} \ \mbox{if\, $-1 \le H \le 0$}    \\ 
                            G \ge H F \, \hbox{and} \, G \le F/H, 
                         & {} \ \mbox{if\, \, $0 \le H \le 1$}  \\
                            G \le H F \, \hbox{and} \, G \ge F/H, 
                         & {} \ \mbox{if\, \, $1 \le H$}
                           \end{array} \right.$ \\
\hline
\end{tabular}
\end{center}

\newpage 
\begin{center}
Table~2 \ :\ 
Forms of the order parameter and their parameter ranges\\
\end{center}
\begin{center}
     \begin{tabular}{| l | l | l |} \hline
      ($D_0^2$, $x$) & $S_0=(d_1, d_2, d_3)$ & the parameter region    \\ \hline
  1)  ($0$, $0$) &   $d_1 =d_2 =d_3 =0$ & $g_D \ge 0$                       \\ 
  2)  ($-\frac{g_D}{2g'_D}$, $0$) & $d_i \neq 0, \, d_{j\neq i}=0$ &
        $g_D < 0, \, h_D < 0$   \\
  3)  ($-\frac{g_D}{2g'_D}$, $x$) & $d_1^2+d_2^2+d_3^2=D_0^2$ &
        $g_D < 0, \, h_D = 0$   \\
  4)  ($-\frac{g_D}{2(g'_D -h_D/3)}$, $\frac{1}{3}$) & $d_1 =d_2 =d_3 \neq 0$ & 
        $g_D < 0, \, 0 < h_D \le 3g'_D$  \\ \hline
     \end{tabular}
\end{center}
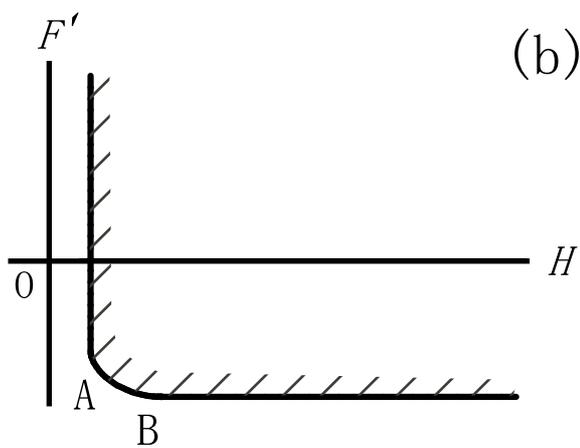
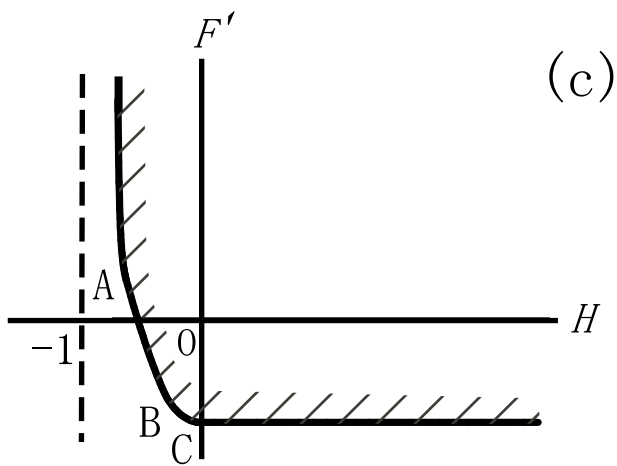
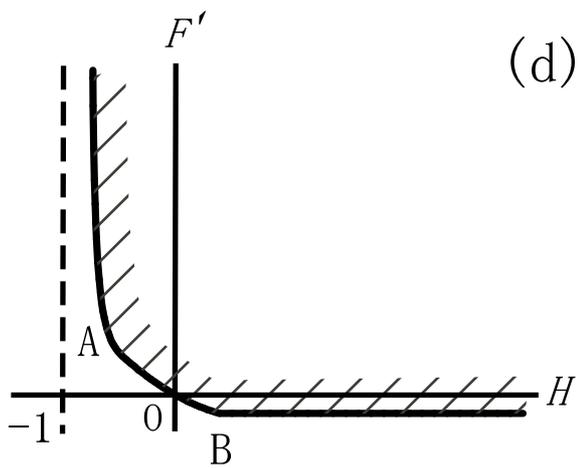
\end{document}